%% file: cioni.tex
\documentclass[graybox]{svmult}

\usepackage{mathptmx}       
\usepackage{helvet}         
\usepackage{courier}        
\usepackage{type1cm}        
\usepackage{makeidx}         
\usepackage{graphicx}        
\usepackage{multicol}        
\usepackage[bottom]{footmisc}

\makeindex             

\begin{document}

\title*{The VMC survey}
\author{Maria-Rosa L. Cioni \& the VMC team}
\institute{Maria-Rosa L. Cioni \at Universit\"{a}t Potsdam, Institut f\"{u}r Physik und Astronomie, Karl-Liebknecht-Str. 24/25, 14476 Potsdam (Germany), Leibnitz-Institut f\"{u}r Astrophysik Potsdam, An der Sternwarte 16, 14482 Potsdam (Germany), and University of Hertfordshire, Physics Astronomy and Mathematics, Hatfield AL10 9AB (United Kingdom); \email{mcioni@aip.de}}
%
%
\maketitle

\texttt{The VISTA survey of the Magellanic Clouds system (VMC) is a public survey project of the European Southern Observatory. It is collecting multi-band near-infrared data across large areas of the Large and Small Magellanic Clouds, the Magellanic Bridge and a few fields in the Magellanic Stream. The combination of great sensitivity to stars below the old main sequence turn off, and the multiplicity at $K_\mathrm{s}$ band (at least $12$ epochs) make VMC highly suitable for the determination of the spatially resolved star formation history and three-dimensional geometry, using variable stars such as Cepheids and RR Lyrae stars. The VMC observations are progressing well and will be completed in 2018. The VMC survey has a high legacy value and many science results have already been published, e.g. the most detailed star formation history map of the Small Magellanic Cloud, and others are in preparation, e.g. a comprehensive investigation of classical Cepheids and a study of the proper motion in the foreground 47 Tuc cluster.}

\section{Survey Status}
\label{sec:1}
The near-infrared $YJK_\mathrm{s}$ survey of the Magellanic Clouds system (VMC; \cite{cio11}; www.\-star.\-herts\-.ac.uk/$\sim$mcioni/vmc)  began in 2009 and will be completed in 2018. It covers about $170$ deg$^2$ in total distributed as follows: $105$ deg$^2$ on the Large Magellanic Cloud (LMC), $42$ deg$^2$ on the Small Magellanic Cloud (SMC), $21$ deg$^2$ on the Magellanic Bridge connecting the two Clouds, and $3$ deg$^2$ on the Magellanic Stream. The sensitivity of the observations allows us to detect sources as faint as the main sequence turn off ($\sim 22$ mag in the Vega scale), sampling the entire age range of stellar populations and allowing us to derive accurate star formation histories. Multi-epochs are collected at each filter: $3$ at $Y$ and $J$, and at least $12$ at $K_\mathrm{s}$. The latter allows us to obtain accurate mean magnitudes for variable stars (e.g. Cepheids and RR Lyrae stars) that are used to measure the three-dimensional geometry of the system. The VMC data have a high legacy value not only for studies of the Magellanic Clouds but also for studies of the Milky Way foreground and of background galaxies (including quasars, \cite{cio13}).

VMC data are obtained from the VISTA telescope in Chile and are processed at both the Cambridge Astronomy Survey Unit (CASU, \cite{irw04}) in Cambridge and at the Wide-Field Astronomy Unit (WFAU, \cite{ham04}) in Edinburgh using the VISTA Data Flow System pipeline (VDFS; \cite{eme04}). CASU performs the standard processing of the individual images, and produces source catalogues at each epoch of observation for each detector (the VISTA camera has sixteen detectors in the field-of-view $=$ pawprint) and for each tile (the mosaic of six pawprints to fill a contiguous area of sky $=$ tile). WFAU links the individual epochs and combines the observations of the same tiles re-extracting sources and producing deep catalogues. We also derive point-spread function photometry to detect fainter sources and improve detections in crowded regions. The VMC data are periodically released to the community via the VISTA Science Archive (\cite{cro12}) and via the ESO archive.

VMC observations are performed in service mode which guarantees a high level of homogeneity of the data. The resulting FWHM of sources in the densest regions is $\le 1^{\prime\prime}$ at each filter while for sources in the outermost regions of the system it is  $0.1^{\prime\prime}-0.2^{\prime\prime}$ higher. Individual tiles correspond to $1^\mathrm{h}$-long observations and reach a sensitivity of $Y=21.2$, $J=21.4$, and $K_\mathrm{s}=19.6$ mag. The overall completion of the survey is $>60\%$: the two Stream tiles are fully observed, the Bridge and SMC are $>80\%$ observed, and the LMC is $\sim50\%$ observed. Five tiles in the LMC, including tiles overlapping the South Ecliptic Pole (SEP) and the 30 Doradus regions, and two tiles in the SMC are already publicly available.

The VMC survey is the only near-infrared survey of the Magellanic system that is complementary to other ongoing large-scale optical surveys: the OGLE-IV survey that will provide the variability information across an unprecedented area of sky, the SMASH survey that searches for stellar debris in the outermost regions of the system, the STEP survey at the VST that focuses on the star formation history of the SMC and Bridge,  Skymapper observations of the southern sky with a few repeats, and Gaia observations that will provide accurate positions and proper motions for all stars down to a magnitude below that of the horizontal branch, including RR Lyrae stars.  The combination between VMC and these surveys will be highly valuable to investigate the origin and evolution of the system.

\section{Science Highlights}
\label{sec:2}

Within the past five years, the VMC team has published $15$ articles in refereed journals that address a variety of science topics and that demonstrate the high quality and versatility of the VMC data. We studied planetary nebulae (\cite{mis11}) and asymptotic giant branch stars (\cite{gul12}), we measured the star formation history (\cite{rub12}) and the parameters of star clusters (\cite{pia14}, \cite{pia15}), we analysed the periodicity of Cepheids (\cite{rip12}, \cite{rip14}, \cite{rip15}) and set the stage for further variability studies (\cite{mor14}), we identified eclipsing binary stars (\cite{mur14}) and quasar candidates (\cite{cio13}), we built a map of the dust distribution (\cite{tat13}), and studied population gradients (\cite{li14}) of Milky Way star clusters. We report below a brief summary of some recent and ongoing investigations.

The most accurate star formation history map of the SMC has been obtained using the VMC data (\cite{rub15}). It was derived from the analysis of 10 tiles covering the main body. Each tile was subdivided into 16 equal regions, and for each region colour-magnitude diagrams were created. These showed a range of stellar populations that sample the entire history of the galaxy. We found, for each diagram, the best combination of 'partial' models (\cite{rub12}) that describes the stellar populations using the STARFISH (\cite{har01}) code. The reddening and the distance corresponding to each region were derived from a range of input values and then the associated star formation rate and iron abundance were analysed as a function of time. Overall, the star formation rate across the SMC was modest at ages $<5$ Gyr $(0.15 M_\odot$yr$^{-1}$ until a major burst that was responsible for the creation of most of the galaxy mass. Another star formation peak, at $1.5$ Gyr, agrees with peaks in the formation of star clusters and in the LMC. The most recent star formation took place in the centre and in the eastern part of the galaxy.

The VMC survey has already covered a large fraction of the LMC providing counterparts for about $40\%$ of the known classical Cepheids (Ripepi et al. in preparation). These stars follow a tight period-luminoisty-colour relation that is a powerful distance indicator. Previous works on smaller samples of classical Cepheids in the SEP and 30 Doradus regions (\cite{rip12}), and on both anomalous and type II Cepheids, indicate dispersions $\le0.1$ mag (\cite{rip14}, \cite{rip15}). The $13$ VMC data points sample well the $K_\mathrm{s}$-band light-curves of Cepheids and for most of the stars the $4$ $J$-band data points are also well representing the expected sinusoidal variation.

RR Lyrae stars, that are about 3.5 magnitude fainter than classical Cepheids, show a lower $K_\mathrm{s}$-band amplitude, but their light-curve is well sampled by the VMC measurements. Their dispersion around the period-luminosity relation is larger and this is probably due to the stars belonging to a spheroid rather than to a disk, like in the case of Cepheids. In the SEP region there are $109$ confirmed RR Lyrae stars with a VMC counterpart ($79$ RRab, $23$ RRc, and $7$ RRd), Moretti et al. (in preparation).


%
\begin{figure}[ht]
\sidecaption
\includegraphics[scale=.5]{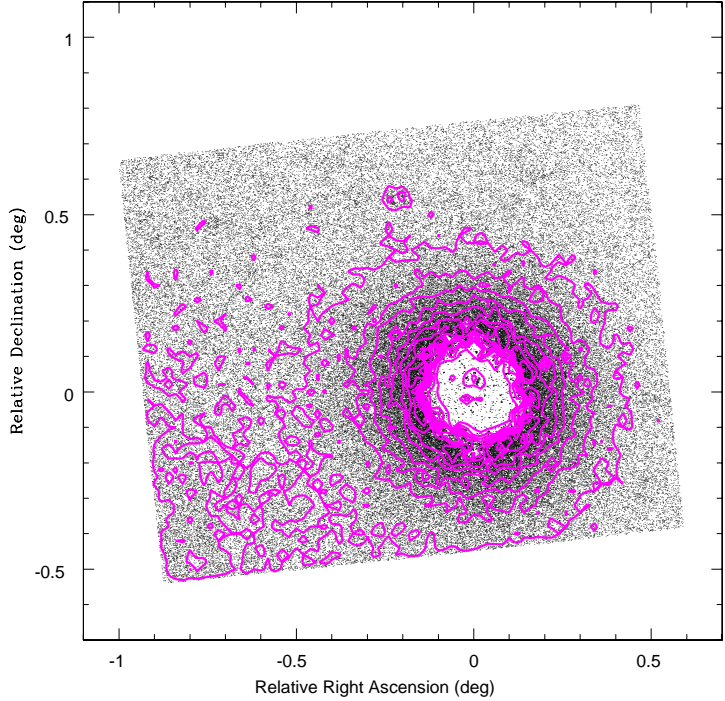}
%
%
\caption{Distribution of VISTA sources over tile SMC $5\_2$. The Milky Way star cluster 47 Tuc occupies about $1/4$ of the area while the SMC star cluster NGC 121 is located to the  North of it. The centre of the figure corresponds to the centre of the 47 Tuc cluster at $(\alpha_0$, $\delta_0)=6^\circ.023625$, $-72^\circ.081283)$. The SMC stellar population density increases to the South-East of 47 Tuc.}
\label{fig:1}       
\end{figure}

Projected onto the SMC background and to the North-West of the galaxy centre is the Milky Way star cluster 47 Tuc (Fig.~\ref{fig:1}). In this tile ($1.5$ deg$^2$ in size) we trace the stellar cluster density out to large radii, detect the increase in the star counts to the direction of the SMC centre and isolate the stars of the SMC star cluster NGC 121. In the colour-magnitude diagram, $(J-K_\mathrm{s})$ vs. $K_\mathrm{s}$, these different stellar populations are easily distinguished from each other and from the large number of background galaxies, as well as from foreground stars. The VMC observations of this tile (SMC $5\_2$) have been used to measure the proper motion of the different stellar populations (Cioni et al. in preparation). We used $10$ separate $K_\mathrm{s}$ epochs obtained over a time-baseline of about one year. The proper motion of individual sources results from a linear fit to the pixel displacements among the different epochs calculated from individual VISTA detectors (sixteen) and paw prints (six) placed on the same reference system defined by background galaxies (there are a few hundreds galaxies per detector). Then, the proper motion for different stellar populations, selected from the colour magnitude diagram, e.g. SMC red clump stars, SMC upper and lower red giant branch stars, Milky Way stars, 47 Tuc upper and lower main sequence stars, and 47 Tuc red giant branch stars are derived by averaging the values for all stars within the given regions applying a sigma clipping method (see Cioni et al. 2014 for details). Proper motion results indicate a clear separation between the SMC, the Milky Way, and the 47 Tuc populations. Furthermore, uncertainties are sufficiently small to distinguish between stellar populations of a different age. Combining all $30,000$ SMC stars and all $80,000$ cluster stars we derive proper motion accuracies of $0.1$ mas yr$^{-1}$ ($25$ km/s). This method is superior to the pilot study described in \cite{cio14} because it is almost free from systematic uncertainties since it is based on the analysis of separate detectors, instead of using coordinates obtained after reconstructing a tile (each detector has its own distortion pattern that is difficult to evaluate after it is combined with the others in the tile mosaic).

\section{Conclusions}
\label{sec:3}
The VMC survey is continuing to observe the Magellanic system acquiring data of unprecedented quality at near-infrared wavelengths. The survey will be completed in 2018 and some data area already publicly available. The accuracy of the VMC data permits the study of different stellar populations, derive their age, metallicity, reddening, distance, and absolute proper motions with respect to a reference system made of background galaxies. The VMC survey has a high legacy value and provides a wealth of targets for future spectroscopic investigations, e.g. with the 4MOST (at VISTA) and MOONs (at the VLT) spectrographs in the 2020s. The VMC survey is a public survey project of ESO and the VMC team welcomes enthusiastic astronomers who would like firsts in the exploitation of this rich data set.

\begin{acknowledgement}
MRLC acknowledges support by the German Academic Exchange Service (DAAD).
\end{acknowledgement}

\input{referenc}

\end{document}

%% file: referenc.tex
%
%
%

%% file: cioni.bbl
\begin{thebibliography}{99.}

\bibitem{cio11} M.-R.L. Cioni, G. Clementini, L. Girardi, et al., 2011, A\&A 527, A116
\bibitem{cio13} M.-R.L. Cioni, D. Kamath, S. Rubele, et al. 2013, A\&A 549, A29
\bibitem{cio14} M.-R.L. Cioni, L. Girardi, M.I. Moretti, et al. 2014, A\&A 562, A32
\bibitem{cro12} N.J.G. Cross, R. S. Collins, R.G. Mann, et al. 2012, A\&A 548, A119
\bibitem{eme04} J.P. Emerson, M.J. Irwin, J. Lewis, et al. 2004, SPIE Conf. Ser. 5493, 401
\bibitem{gul12} M. Gullieuszik, M.A.T. Groenewegen, M.-R.L. Cioni, et al. 2012, A\&A 537, A105
\bibitem{ham04} N.C. Hambly, R.G. Mann, I. Bond, et al. 2004, SPIE Conf. Ser. 5493, 423
\bibitem{har01} J. Harris, D. Zaritsky 2001, ApJS 136,  25
\bibitem{irw04} M.J. Irwin, S. Hodgkin, et al. 2004, SPIE Conf. Serr. 5493, 411 
\bibitem{li14} C. Li, R. de Grijs, L. Deng, et al. 2014, ApJ 790, 35
\bibitem{mis11} B. Miszalski, R. Napiwotzki, M.-R.L. Cioni, et al. 2011, A\&A, 531, A157
\bibitem{mor14} M.I. Moretti, G. Clementini, T. Muraveva, et al. 2014, MNRAS 437, 2702
\bibitem{mur14} T. Muraveva, G. Clementini, C. Maceroni, et al. 2014, MNRAS 443, 432
\bibitem{pia14} A. Piatti, R. Guandalini, V. Ivanov, et al. 2014, A\&A 570, A74
\bibitem{pia15} A. Piatti, R. de Grijs, S. Rubele, et al. 2015, MNRAS accepted
\bibitem{rip12} V. Ripepi, M.I. Moretti, M. Marconi, et al. 2012, MNRAS 424, 1807
\bibitem{rip14} V. Ripepi, M. Marconi, M.I. Moretti, et al. 2014, MNRAS 437, 2307
\bibitem{rip15} V. Ripepi, M.I. Moretti, M. Marconi, et al. 2015, MNRAS 446, 3034
\bibitem{rub15} S. Rubele, L. Girardi, L. Kerber, et al. 2015, MNRAS in press
\bibitem{rub12} S. Rubele, L. Kerber, L. Girardi, et al. 2012, A\&A 537, 106
\bibitem{tat13} B. Tatton, J.Th. van Loon, M.-R.L. Cioni, et al. 2013, A\&A 554, 33


%
%
%
\end{thebibliography}
